\documentclass[runningheads]{llncs}
\usepackage[T1]{fontenc}
\usepackage{graphicx}
\usepackage{multirow}
\usepackage{balance} 
\usepackage{stfloats}
\usepackage{xcolor}
\usepackage{booktabs}
\usepackage[most]{tcolorbox}
\usepackage{url}
\usepackage{xcolor}
\usepackage{comment}
\usepackage{subcaption}

\begin{document}
\title{“We Used It Anyway”: Low Trust and High Adoption of Large Language Models in Requirements Engineering Education}
\titlerunning{Low Trust and High Adoption of LLMs in RE Education}
%
%
\author{
  \IEEEauthorblockN{Riccardo Coppola, Giacomo Garaccione, Stefano Mancini, Luca Ardito}
  \IEEEauthorblockA{Politecnico di Torino, Italy\\first.last@polito.it}
}

\author{Riccardo Coppola\inst{1}\orcidID{0000-0003-4601-7425} \and
Giacomo Garaccione\inst{1}\orcidID{0000-0001-7254-9578} \and
Stefano Mancini\inst{1} \and
Luca Ardito\inst{1}\orcidID{0000-0002-0501-7886}}
\authorrunning{R. Coppola et al.}
%
\institute{Politecnico di Torino, Torino, Turin, Italy 
\email{first.last@polito.it}}
\maketitle              
\begin{abstract}
Recent advances in Large Language Models (LLMs) are reshaping how students approach Software Engineering activities, particularly those centered on textual artifacts. This paper reports on the experience of integrating a non-restrictive LLM use within a first-year Master’s course in Software Engineering, in which students completed a semester-long Software Engineering project divided in a Requirements Engineering (RE) phase and a software testing phase. Following the completion of the RE phase of the project, we administered a 12-question survey to 43 students to investigate which LLM tools they used, for which activities, how they evaluated AI-generated outputs, and how they envisioned the role of LLMs in future Software Engineering curricula. Our findings show that LLM use was nearly universal among respondents and concentrated primarily on text-intensive activities such as use-case narratives, consistency checking, brainstorming, and refinement of existing artifacts. Despite this widespread adoption, most students perceived AI-generated outputs as inferior to their own work, frequently reporting issues related to incorrect or incomplete content, mismatches with course-specific conventions, over-engineering, and limitations in diagram generation. Nevertheless, students continued to rely extensively on LLM as a reviewer, critic, and productivity aid rather than as a fully autonomous artifact generator. We also observe that students’ choice of models was strongly influenced by access to premium subscriptions available for free for student accounts, highlighting the role of economic accessibility in shaping educational AI ecosystems. The results suggest that the most significant challenge in the adoption of LLMs within Requirements Engineering education is not whether students use these tools, but whether they possess the skills required to critically evaluate, contextualize, and correct its outputs. 

\keywords{Generative AI \and Large Language Models \and Requirements Engineering \and Software Engineering Education \and Project-based learning \and Experience reports}
\end{abstract}
%
%
\section{Introduction}
\label{sec:intro}

Recent years have witnessed the emergence and diffusion of freely accessible Large Language Models (LLMs), impacting significantly how students work in higher education. These systems can, in fact, produce extended and plausible textual artifacts on most topics and for most of the assignments that are administered to the students. In Software Engineering courses, many Requirements Engineering (RE) assignments are fundamentally textual and require the production of largely natural language-based documents (e.g., requirements documents, use case narratives, glossaries). They involve the translation of domain knowledge and stakeholder intent into structured textual and graphical artifacts. making them very plausible targets for LLM assistance \cite{camara2023assessment} \cite{wang2024llms_uml}. However, empirical studies have also shown that LLMs can be an effective support tool for text-intensive tasks such as brainstorming or drafting, while at the same time presenting relevant issues such as lack of domain understanding, inconsistencies, and a tendency to hallucinate when asked to generate graphical, context-dependent artifacts \cite{camara2023assessment}. RE activities are highly context-dependent, and a requirements document (for both real and simulated systems) must reflect stakeholder perspectives, constraints, and conventions that might be unknown to a general-purpose LLM. At the same time, LLMs remain a tool that future Software Engineers must properly understand, being aware of the risks coming from a superficial utilization; it is of paramount importance to integrate traditional SE courses with guidance on how to use LLMs to assist SE tasks. Despite the rapid growth of empirical research on LLMs in computer science and engineering education in general~\cite{axler2025assessing} \cite{hou2025evolving}, few works have examined how students use them \emph{within} a structured, multi-deliverable software engineering project. This paper presents an experience report gathered within a master's course in Software Engineering, through the administration of a survey to 43 students who had completed the requirements engineering phase of a semester-long group project. The survey aimed at collecting data on tool choice, task coverage, supported activities, output quality, model comparison strategies, and proposal for future AI integration in the course. The goal of this study is not to mandate a policy on LLM utilization or adoption in Software Engineering courses, but rather to interpret the natural utilization of LLMs that emerged spontaneously among students when tasked with traditional requirements engineering tasks. The final objective of the analysis is to gather and report actionable recommendation for curriculum and project design for SE courses. The remainder of this paper is organized as follows. Section~\ref{sec:background} reviews related work on LLM adoption in engineering education and on LLMs in requirements engineering. Section~\ref{sec:course} describes the course, the project structure, and the survey design. Section~\ref{sec:results} presents the results, which we discuss in Section \ref{sec:disc}. The lessons learned from the study are presented in Section~\ref{sec:lessons}, and Section~\ref{sec:threats} analyzes the threats to the study's validity. Lastly, Section~\ref{sec:conclusion} draws conclusions and outlines the next steps we plan to undertake following the survey.

\section{Background and Related Work}
\label{sec:background}
Prior work has extensively explored students’ perceptions of LLMs across generic higher education contexts, highlighting a combination of strong perceived utility and persistent concerns regarding ethics, trust, and educational impact. Noroozi et al. investigated students’ perceptions of LLMs in the context of peer feedback activities \cite{noroozi2025value}, finding that while only 44\% of students actively used LLMs, most participants still perceived the technology as moderately or very helpful for low-level writing tasks; concerns were however expressed regarding inaccurate information, unreliable citations, and potential reduction of learning opportunities. Similarly, Maxwell et al. examined demographic differences in student perceptions of LLMs across a large-scale survey involving 482 higher education students \cite{maxwell2025generative}, revealing widespread adoption and high levels of comfort with LLMs for both general and academic purposes, with reported improvements in productivity, support to academic success, albeit with similar concerns to those raised by the work by Noroozi et al., while also highlighting demographic differences impacting LLM perception. Another relevant contribution is provided by Chan and Zhou, who developed an Expectancy-Value Theory (EVT)-based instrument to investigate how students’ perceptions influence their intention to use LLMs in higher education \cite{chan2023deconstructing}, finding a strong positive correlation between perceived LLM value and willingness to use such a tool for academic purposes, together with perceived drawbacks such as ethical concerns and potential negative consequences negatively impacting intention to use LLMs. Daher et al. explored higher education students’ perceptions of LLM tools across cognitive, social, affective, and behavioral dimensions \cite{daher2024higher}, and found that students described LLM systems as helpful, motivating, and capable of increasing their confidence.

\subsection{LLM Adoption in Requirements Engineering Education}
\label{sec:bg-adoption}

Narrowing the focus to Software and Requirements Engineering (RE), using Large Language Models in education introduces both promising opportunities and important challenges. Students and educators are increasingly exploring how tools like ChatGPT can support the teaching and learning of core RE competencies, and these tools have demonstrated particular value in enhancing brainstorming activities and helping students understand complex problems from multiple perspectives, thereby providing more realistic views of underlying complexities in real-world requirements engineering tasks \cite{tiwari2025leveraging}. However, the adoption of LLMs in RE education requires careful pedagogical design to maintain learning integrity and skill development. Research shows that while students appreciate AI's efficiency in generating requirements and its ability to accelerate development processes, they simultaneously recognize the critical need for human oversight to ensure accuracy and quality \cite{mellqvist2024exploring}. A hybrid approach that combines AI-generated and manually crafted requirements has proven most effective, balancing artificial intelligence's advantages with essential human insights and domain expertise. Additionally, educators must address practical challenges such as students' difficulties in creating effective prompts and the time-consuming process of refining AI-generated outputs to meet curriculum standards. To maximize educational value, institutions implementing generative AI in RE courses should establish clear frameworks for responsible tool use, emphasizing how LLMs can serve as a pedagogical partner that enhances rather than replaces the critical thinking and analytical skills that define professional requirements engineering practice \cite{ayora2025rethinking}.

\subsection{LLMs in Requirements Engineering}
\label{sec:bg-llmre}

Extensive research has been performed regarding requirements engineering tasks that can be automated with LLMs. Hemmat et al. reviewed the application of Large Language Models in Software Requirements Engineering  \cite{hemmat2025research}, showing that LLMs are increasingly employed in requirements elicitation, specification, modeling, validation, and quality assurance tasks such as generating UML diagrams, refining requirements, supporting traceability, and automating consistency checking. The paper also analyses optimization strategies such as prompt engineering, fine-tuning, and LoRA adaptation, while highlighting challenges related to hallucinations, poor domain understanding, and structural inconsistencies in generated artifacts. Arora et al. analyzed the role of LLMs across the complete requirements engineering lifecycle \cite{arora2024advancing}, discussing how they can support elicitation, specification, and analysis; additional focus is also given to validation activities, while also warning that current LLMs remain limited by contextual understanding, explainability issues, hallucinated outputs, and difficulties in handling complex project-specific constraints. Khan et al. identified elicitation, modeling, prioritization, specification, validation, and traceability as the most common application domains for LLMs in Requirements Engineering\cite{khan2025large} and reported that GPT-based systems dominate the current literature because of their effectiveness in processing natural language artifacts and generating structured documentation, while also discussing common prompting strategies, evaluation methodologies, and limitations such prompt sensitivity, lack of standardized benchmarks, and difficulties in evaluating the semantic correctness of generated requirements artifacts. A specific research thread focused on LLMs for RE activities (LLM4RE) concerns the automated generation of graphical modeling artifacts starting from natural language descriptions. Cámara et al. \cite{camara2023assessment} performed one of the earliest studies on this trend, exploring the capabilities of ChatGPT when applied to Unified Modeling Language (UML) tasks, finding that the tool struggled on multiple sides, with syntactic deficiencies, inconsistencies in the response format, and scalability limitations when producing structural models. Ferrari et al. \cite{ferrari2024model} also investigated ChatGPT's ability to generate diagrams, focusing on UML sequence diagrams, and identified key challenges such as needing iterative, RE-focused prompting strategies and requiring a way to express tacit domain knowledge when communicating with agents. Similar conclusions were reached by Wang et al. \cite{wang2024llms_uml}, who found that LLMs can be effective assistants, while at the same time presenting limitations that require careful human intervention. Our prior work also explored this research area, exploring the ability of LLM agents to generate UML class diagrams in RE-oriented modeling exercises, proving that LLM agents have a good understanding of modeling rules and conventions, while at the same time presenting significantly worse ability to represent domain constraints \cite{debari2024uml}; a subsequent study compared multiple LLM agents and observed that certain LLMs are more suited for different quality aspects, and that detailed prompting strategies can improve diagram quality \cite{garaccione2025comparison}.
 
\section{Methodology}
\label{sec:course}

In this section we document the structure of the course and the course-long project, and the survey that was administered to the students.

\subsection{Course outline}
 
The study was conducted in the context of a Software Engineering course offered in the first year of a Master's degree programme in Computer Engineering at the Polytechnic University of Turin. The course is mandatory for students in the Computer Engineering orientation and is taken in the second semester of the programme. The background of the students is typically a bachelor's degree in Computer Engineering or closely related fields, but with limited experience in professional software development projects. The course is organized around lectures on SE principles and is paired with a non-mandatory course-long project, which is evaluated for 10 points out of the total grade's 33. The number of enrolled students in the course ranges between 300 and 400 yearly, divided into two separate editions of the course, taught respectively in Italian and English. The course's topics aim to provide a general overview of the software engineering pipeline, starting from requirements engineering activities in its first half before moving to testing techniques in the latter part; both halves see a combination of theory and practice introduced through the creation of an example platform for music sharing: in the first half of the course, when a new topic is introduced during theory lectures, the corresponding part of the requirement document is produced in practical sessions. In the second half, the teachers provide an implementation of the platform that is then used to conduct live examples of how to apply different testing techniques. 

\subsection{Project Structure}

The course project consisted of the design and specification of Participium, a citizen-participation web platform for supporting the reporting and management of urban issues (geo-localized reports, map/table views, lifecycle management, notifications, and administrative dashboards). The system was conceived as an open-source platform potentially reusable by other municipalities, requiring students to reason about both citizen-facing interaction workflows and internal operational processes of public administrations. The course project is divided into two sequential phases. The first phase, covering roughly the first 6 weeks of the course, covers the activities related to the Requirements Engineering lifecycle, which are referred to as \emph{tasks}: I) \textbf{Task 1: Project Management:} Product Breakdown Structure (PBS), Work Breakdown Structure (WBS), Gantt chart, and risk register; II) \textbf{Task 2: Requirements Engineering:} Stakeholder analysis, context diagram, user personas, user stories, and functional/non-functional requirements; III) \textbf{Task 3: Use Cases:} Use-case diagram, narrative descriptions for each use case, and a traceability table linking use cases to functional requirements; IV) \textbf{Task 4: Architecture and Design:} Class diagram and deployment diagram. The description of the Participium platform given to the students at the beginning of Phase 1 as the context for the project is available online, together with the reference solution for the first phase, provided to the students after the deadline for the submission of their deliverables\footnote{\url{https://doi.org/10.6084/m9.figshare.32287281}}. The second phase is instead split into six assignments that introduce testing concepts by asking students to define test suites on a demo platform developed by course instructors, with a new assignment each week and a platform update, to simulate a real-world scenario where students act as testers for an application developed by other developers.

\subsection{Group composition}
 
Groups of 4--5 students work on a common software project specification throughout all phases of the project. The groups are constructed algorithmically in order to maximize intra-group and inter-group diversity, by adopting the algorithm described by Torchiano et al. \cite{torchiano2025impact}. Each group of students is assigned a private repository hosted on the versioning system used in the course (namely, GitLab). All task submissions are done by pushing to the main branch of the repository, which is forked by a master repository curated by the instructors, where the instructions for each task are pushed weekly. Task submissions are then reviewed by the instructors and returned along with formative feedback (in the form of textual comments on the deliverables) before the next task begins.

\subsection{Survey Design and Analysis}
\label{sec:survey}
 
The survey comprised 12 questions, a mix of closed-ended (single-choice and multiple-choice) and open-ended items, grouped into thematic blocks related to the following Research Questions (RQs): 

\begin{itemize}
    \item \textbf{RQ1:} Which models are used the most by students?
    \item \textbf{RQ2:} In which tasks are LLMs used the most?
    \item \textbf{RQ3:} What is the perceived quality of the LLM agents' output?
    \item \textbf{RQ4:} How do students envision the integration of LLMs in SE curricula?
\end{itemize}

The full questionnaire is reported in Table \ref{tab:questionnaire}.

\begin{table*}
    \small
    \caption{Questionnaire administered to the students}
    \resizebox{\linewidth}{!}{%
    \begin{tabular}{lp{12cm}l}
    \toprule
    Question ID & Question & Question type\\
    \midrule
    1.1 & Which LLM tools did you use in the project task? & Multiple choice\\
    1.2 & Which tool did you prefer and why? (in case of multiple selection in 1.1) & Open\\
    \midrule
    2.1 & In which tasks did you use LLM tools? & Multiple choice\\
    2.2 & In which task did you use LLMs the most? & Single choice\\
    2.3 & Which was the motivation for using LLMs in that task? & Open\\
    2.4 & Which activities were supported the most by the use of LLMs? & Multiple choice\\
    \midrule
    3.1 & Was the output of AI perceived as worse than your own work? & Single choice\\
    3.2 & Which deliverables required significant correction, and why? & Open\\
    3.3 & Did you compare the output from multiple models? & Single choice\\
    3.4 & How did you select the output when comparing multiple models? & Open\\
    \midrule
    4.1 & Do you believe that LLMs can be usefully integrated in the course? & Single choice\\
    4.2 & How could such integration work? & Open\\
    \bottomrule
    \end{tabular}%
    }
    \label{tab:questionnaire}
\end{table*}

The survey was hosted on the Google Form platform and administered in English to both editions of the course. Open-ended responses were analyzed using open coding, a qualitative method rooted in grounded theory in which codes are assigned inductively from the data, grouped into themes, and then revised through a second pass to improve consistency~\cite{strauss1990basics} \cite{treude2024qualitative}. The open-coding procedure was performed by one of the authors of the manuscript, with a second pass involving all authors for checking the assignment of codes. Each open-ended question was coded independently, using a per-question theme set; a response could be assigned only one theme per question. Closed-ended responses were summarized by frequency counts.

\section{Results and Discussion}
\label{sec:results}

In this section, we report the results of the questionnaire administered to the course students. The survey received 43 respondents (28 male, 14 female, 1 non-binary/prefer not to say), with ages ranging from 21 to 32 (median 23).

\subsection{RQ1: Tool Adoption and Preference}
\label{sec:res-tools}
 
\begin{figure}[t]
  \centering
  \begin{subfigure}[t]{0.48\linewidth}
    \centering
    \includegraphics[width=\linewidth]{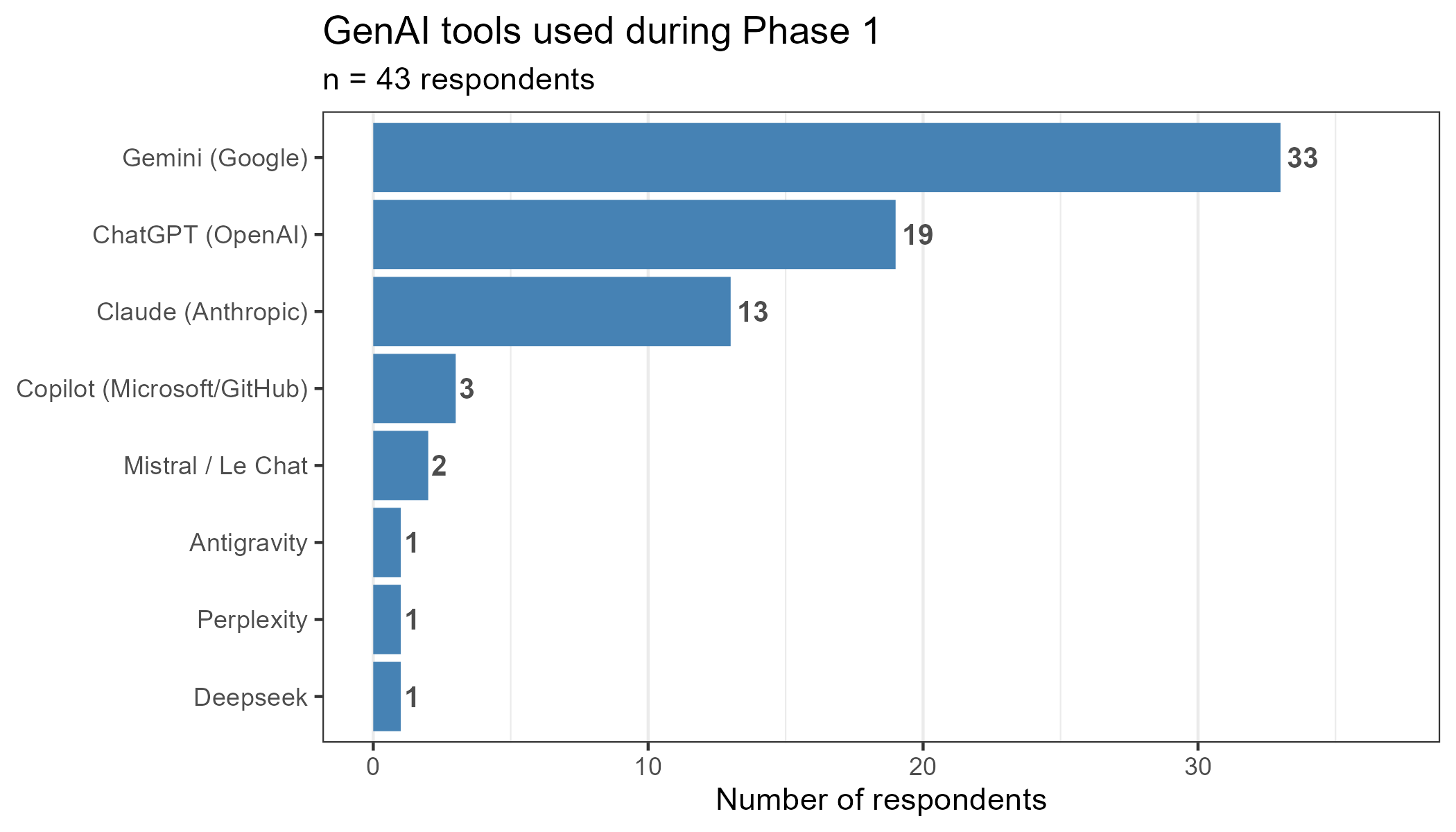}
    \caption{Question 1.1: LLM tools used in the project}
    \label{fig:rq1-adoption}
  \end{subfigure}
  \hfill
  \begin{subfigure}[t]{0.48\linewidth}
    \centering
    \includegraphics[width=\linewidth]{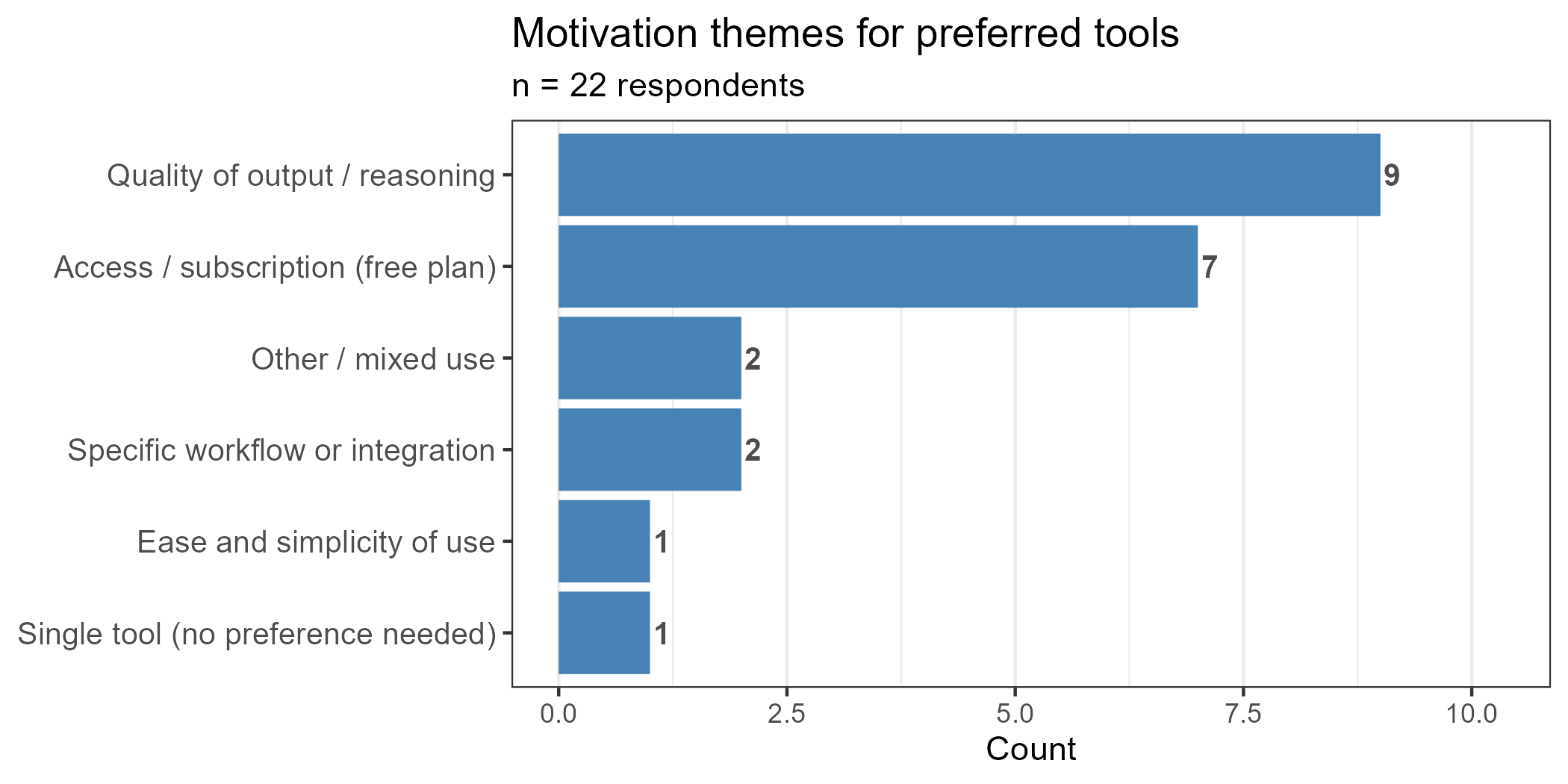}
    \caption{Question 1.2: motivations behind the preference of a specific LLM}
    \label{fig:motivation}
  \end{subfigure}
\end{figure}

LLM use during Phase 1 was essentially universal: all 43 respondents reported using at least one tool. In Figure. \ref{fig:rq1-adoption} we report the adoption of different tools. Gemini (Google) was by far the most widely used tool (33 respondents, 77\%), followed by ChatGPT/OpenAI (19, 44\%) and Claude by Anthropic (13, 30\%). Smaller numbers reported using Copilot (3), Mistral/Le Chat (2), Perplexity (1), and Deepseek (1). These counts allow for multiple tools per respondent.

Figure \ref{fig:motivation} shows the frequencies of the codes applied to the answers of question 1.2. 7 of 22 respondents who used multiple tools mentioned that the decisive factor was the availability of a Pro subscription plan offered free to student accounts, with several explicitly noting that they would have preferred Claude or another tool had access been equivalent. This finding aligns with prior observations that students' tool choices in naturalistic settings are strongly conditioned by economic constraints~\cite{axler2025assessing}. Only 10 of the 22 respondents cited quality of reasoning or output as their primary reason for preferring one tool over another. Notably, Claude was cited most frequently when quality was the stated criterion: respondents who preferred it described it as \emph{less dispersive}, \emph{more critical}, and less prone to simply agreeing with whatever the user proposed. Two respondents valued different tools for different sub-tasks, using one for ideation and another for final consistency checks, adopting a pattern of \emph{tool orchestration} that suggests a more sophisticated appropriation of LLM capabilities. \textbf{Answer to RQ1:} The most widely adopted LLM tool by students was Gemini by Google (77\% of the respondents). Tool selection is mostly conditioned by perceived quality of the LLM outputs and economic reasons (i.e., availability of free student licenses).

\subsection{RQ2: Task and Activity coverage}
\label{sec:res-tasks}
 
Table~\ref{tab:tasks} reports the distribution of AI use across the four project tasks. LLM use was not uniform: it was largely used in Task 3 (Use Cases) (36/43, 84\%), which was also identified as the task with the highest LLM use (22/40, 55\%). In Task 2 (Requirements Engineering), LLM agents were used by 32 out of 43 students (74\%), followed by Task 1 (25/43, 58\%) and Task 4 (18/43, 42\%).

\begin{table}[t]
\small
\caption{Distribution of LLM use across project tasks.}
\label{tab:tasks}
\centering
\begin{tabular}{lcc}
\toprule
\textbf{Task} & \textbf{Used Tool (\%)} & \textbf{Most-used (\%)} \\
\midrule
Task 1 (Project Management)   & 25 (58\%) &  8 (20\%) \\
Task 2 (Requirements Eng.)    & 32 (74\%) &  9 (23\%) \\
Task 3 (Use Cases)            & 36 (84\%) & 22 (55\%) \\
Task 4 (Architecture/Design)  & 18 (42\%) &  1 (2\%)  \\
\midrule
Not specified                   & ---       &  3 (8\%)  \\
\bottomrule
\end{tabular}
\end{table}
 
The open coding of question 2.3 reveals the predominance of LLM use in Task 3, which required students to define multiple use cases and write a structured description covering involved actors and related stakeholders, goals, preconditions, primary scenario and exceptional flows, and post-conditions. The large amount of text, the structured format, and the importance of consistency between the graphical representation and the narrative all influenced AI use in this task. The most common motivation (11 respondents) was \emph{consistency and completeness checking}, for goals such as verifying omissions or ensuring consistency with earlier deliverables. The second most common theme (10 respondents), the high \emph{volume of text to produce}, also points out to the fact that students prefer using LLMs to automate tedious and wordy activities. Other, less common, motivations such as \emph{brainstorming and idea generation} (6) and \emph{lack of domain or contextual knowledge} (5) indicate that students interact with LLMs to enumerate details that they may have missed, to compensate for the unfamiliarity with real-world software development practices that the project implicitly required, or to provide contextual information connected with the municipality-related aspects of the project. This use of LLMs as a proxy for professional experience reflects a legitimate pedagogical gap in courses that ask students to reason about industrial software systems without giving them adequate domain understanding. At the same time, it poses the risk of students trying to acquire domain knowledge that could not be present in the training set of general-purpose LLMs, leading in the worst case to hallucinated responses.

\begin{figure}[t]
  \centering
  \begin{subfigure}[t]{0.48\linewidth}
    \centering
    \includegraphics[width=\linewidth]{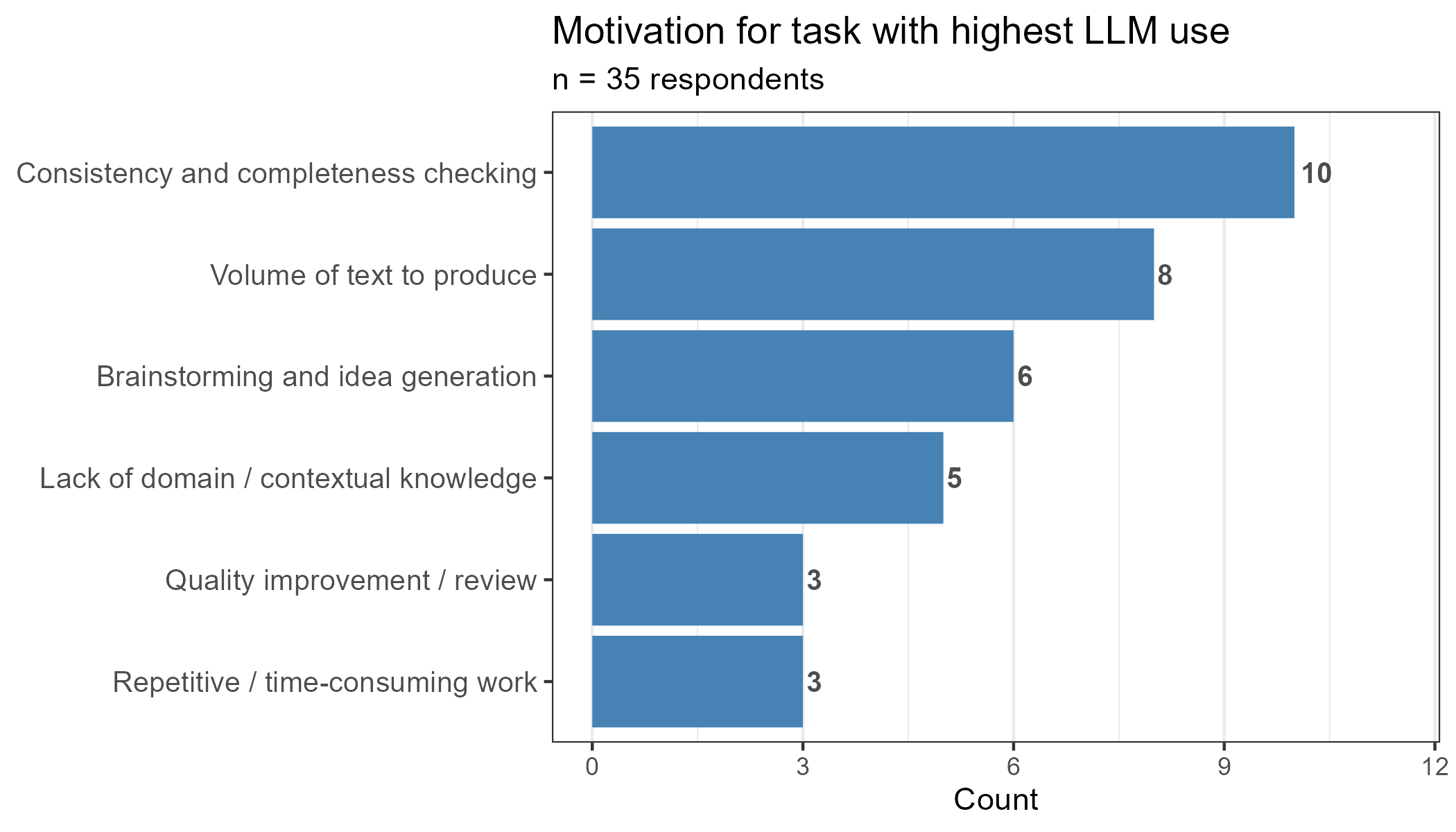}
    \caption{Question 2.3: motivation themes for the task in which AI was used the most}
    \label{fig:motivation2}
  \end{subfigure}
  \hfill
  \begin{subfigure}[t]{0.48\linewidth}
    \centering
    \includegraphics[width=\linewidth]{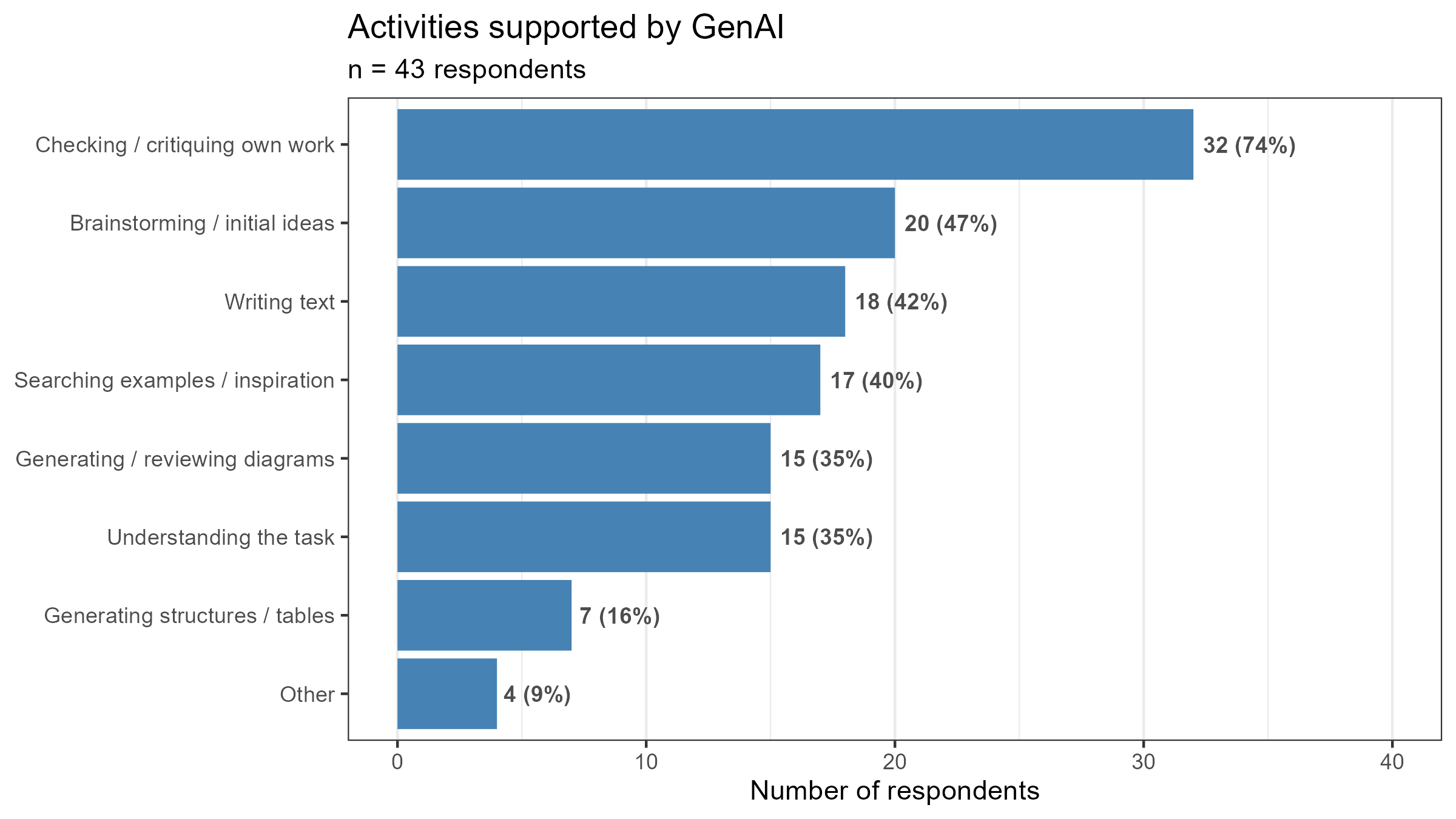}
    \caption{Question 2.4: activities supported by LLMs}
    \label{fig:activities}
  \end{subfigure}
\end{figure}

Question 2.4 allowed multiple selections from a predefined list of alternatives. The results (Figure~\ref{fig:activities}) reveal a strong concentration on \emph{Checking/critiquing own work} (32 mentions, 74\%), indicating that students preferred embedding LLMs in their work pipelines to check the outcome of their autonomous work instead of using them to generate deliverables from scratch. This trend is confirmed by the high number of mentions for \emph{Brainstorming/initial ideas} (29, 47\%),  \emph{Searching examples/inspiration} (17, 40\%), and \emph{Understanding the task} (15, 35\%). The emerging picture shows workflows that are better aligned with educational goals than pure delegation by preserving cognitive engagement with the tasks. \textbf{Answer to RQ2:} The answers to the survey reveal that LLM tools were mostly used in tasks requiring text-intensive and repetitive activities. The most mentioned reasons for using LLMs were checking the students' own work and brainstorming initial ideas, suggesting the use of such tools as reviewers rather than deliverable generators.

\subsection{RQ3: Perceived quality of the LLM outputs}
\label{sec:res-quality}

Question 3.1 was defined to understand whether the students perceived LLM output as worse than their own work. 34 of 43 (79\%) reported that AI-generated output was worse than what they would have produced themselves, while only 4 (9\%) reported the opposite. This finding is non-trivial: despite a widespread perception that AI output was inferior, the students still made extensive use of the LLM tools. This finding pairs well with the utilization of LLMs as checkers and generators of first ideas to be later refined. Perceived inferiority of the LLM output, in fact, does not necessarily translate to deliverable inferiority if the students invest their time in correction after a first draft generation. The open coding of question 3.2 reveals five distinct failure modes, summarized in Figure 5. The most frequent, \emph{incorrect or incomplete content} (9 mentions out of 27 valid responses), covers cases in which AI output was factually wrong, hallucinated, or missing key project-specific elements. Several respondents described AI suggesting actor-to-use-case connections that violated the conventions taught in the course, or generating user stories for features that were not part of the defined project scope.

The second most frequent failure mode, \emph{style and formatting mismatch} (7 mentions), is analytically distinct from content incorrectness: in this case, the LLM output was not semantically wrong but did not conform to the specific format, naming conventions, or structural templates that the course prescribed. This mismatch can only be addressed with extensive prompt engineering (e.g., few-shot prompting by providing examples, providing detailed descriptions) that introduces the course-specific conventions; in fact, multiple students reported that the AI tools were able to follow instructions correctly when supplied with explicit requests and format specifications. \emph{Excessive detail or over-engineering} (4 mentions) and \emph{diagram generation limitations} (4 mentions), the following failure modes, reflect the tendency of LLMs to produce artifacts that are more complex than required, requesting selective reduction of the generated outputs, as well as the fundamental limitation of current LLMs in producing correct, immediately usable diagram representations: even structured formats such as PlantUML or JSON often required significant manual correction. This is consistent with the finding in the broader LLM4RE literature that diagram generation remains one of the weakest capabilities of current agents~\cite{zadenoori2025llmre}.

Slightly more than half of respondents (22/43, 51\%) compared the output of two or more models for at least some tasks; Figure 6 shows the frequencies of codes for the answer to question 3.4. \emph{Personal judgment based on course knowledge} was by far the dominant selection strategy (11/20 respondents who provided a substantive answer), showing that most students did not accept AI output passively but applied their own understanding to evaluate it; it also implies that the quality of AI-model selection is bounded by the student's own conceptual understanding. The secondary strategies, \emph{quality comparison between models} (5 respondents) and \emph{manual merging of outputs} (3 respondents), suggest that some students adopted more sophisticated multi-model workflows: using one model's output to critique or correct another's (a practice similar to the paradigm defined in literature as \emph{llm-as-a-judge} \cite{zheng2023judging}), or selectively combining the best elements of each. \textbf{Answer to RQ3:} The majority of students perceived that the quality of the output generated by LLMs was inferior to their own efforts. The most frequent motivations for subpar output by LLMs were incorrect or incomplete content and errors in style and formatting. Slightly more than half of the respondents performed comparisons between the outputs of different models, both based on personal judgment and using other LLMs as judges.
\begin{figure}[t]
  \centering
  \begin{subfigure}[t]{0.48\linewidth}
    \centering
    \includegraphics[width=\linewidth]{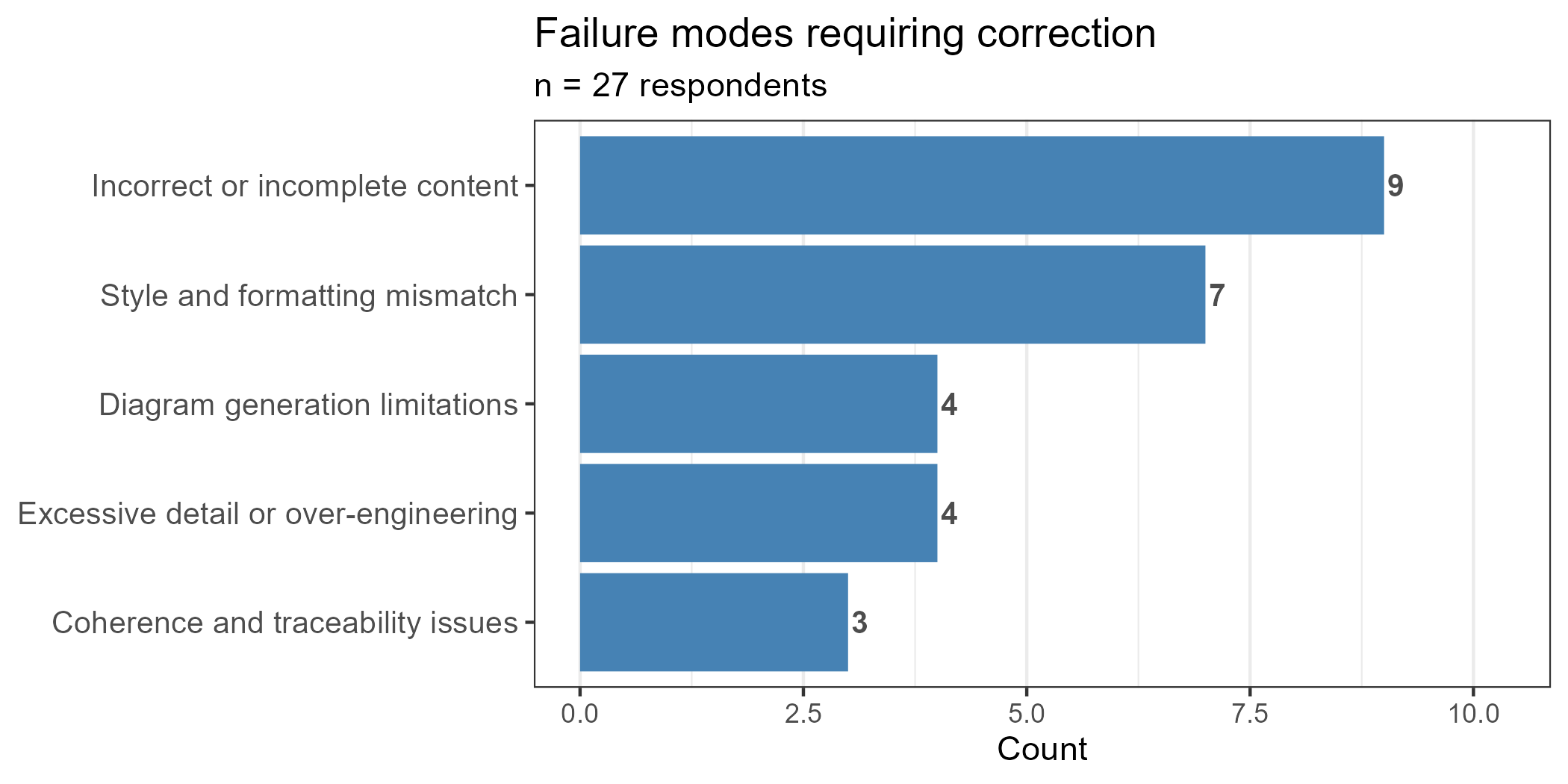}
    \caption{Question 3.2: AI output failure modes requiring correction}
    \label{fig:failmodes}
  \end{subfigure}
  \hfill
  \begin{subfigure}[t]{0.48\linewidth}
    \centering
    \includegraphics[width=\linewidth]{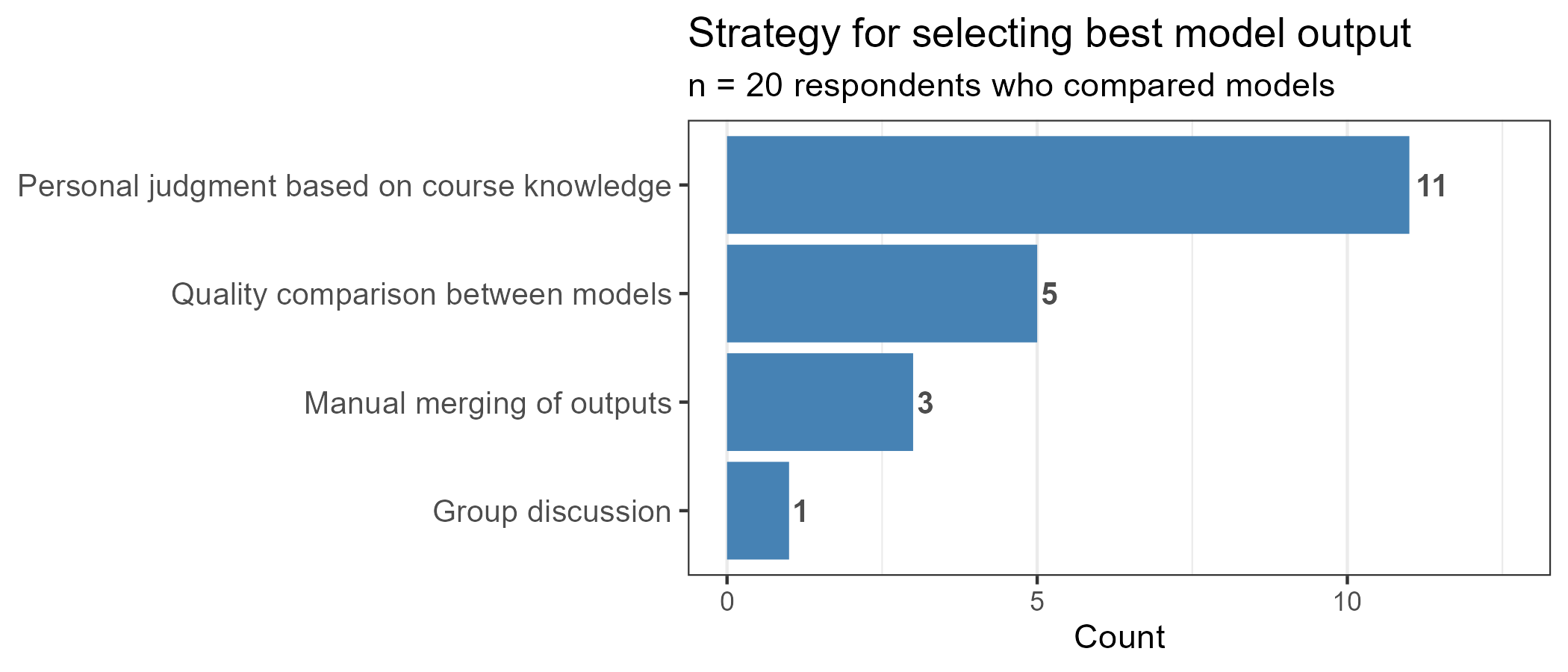}
    \caption{Question 3.4: strategies for selecting the best output when comparing different agents}
    \label{fig:model_sel}
  \end{subfigure}
\end{figure}

\subsection{RQ4: Integration in the Course Curriculum}
\label{sec:res-integration}
 
\begin{figure}[t]
\centering
\includegraphics[width=0.6\linewidth]{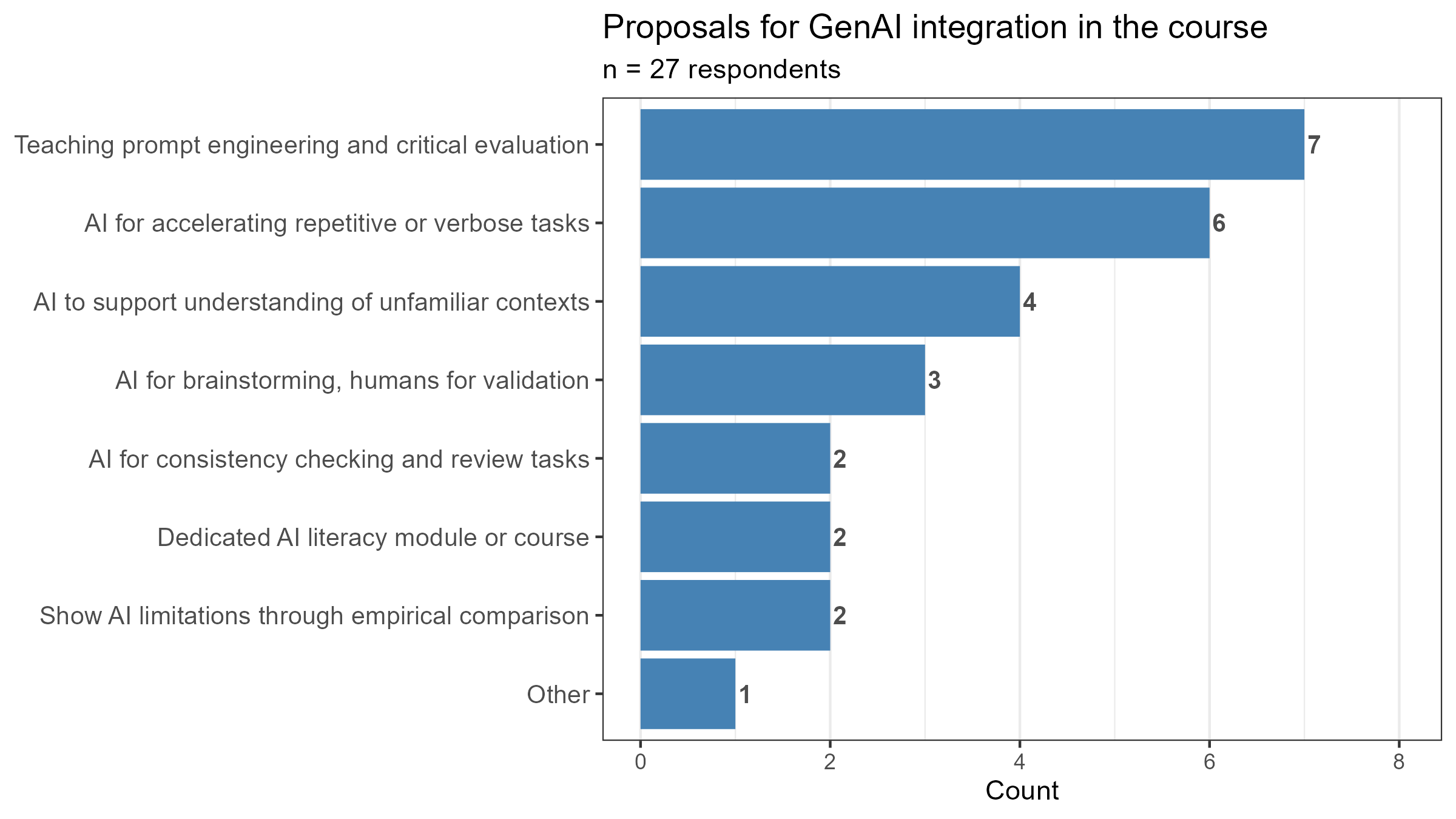}
\caption{Question 4.2: proposed modes of LLM integration in the course}
\label{fig:integration}
\end{figure}

A large majority of respondents (33/43, 77\%) believed that LLMs could be usefully integrated into the course, with only 6 (14\%) disagreeing and 4 (9\%) unsure. The open coding of question 4.2 produced seven themes, whose frequencies are shown in Figure \ref{fig:integration}.

The most frequent theme in the answers is the requirement to teach prompt engineering and critical evaluation of AI output (7 respondents). This answer can be considered as a direct consequence of the failure modes mentioned previously. Students recognize that obtaining useful AI output in a course-specific context requires skills that they currently lack and that the course does not teach. This answer entails important consequences for the management of curricula at higher education institutions, since prompt engineering is a topic of advanced courses in the second year of the master's programme, but it might be beneficial to cover the topic before any Software Engineering course. This is also a valuable insight in the usual debate about the use of LLMs in education: rather than arguing about whether AI should be permitted or not, students are requesting that the curriculum help them use it more effectively and more critically. The second theme is using AI to accelerate repetitive or verbose tasks (6 respondents), in line with the dominant motivation pattern identified in the answer to RQ2. Students view the text-generation capability of AI as a legitimate tool for managing the workload of documentation-intensive tasks, provided the output is carefully reviewed. The theme of \emph{AI to support understanding of unfamiliar contexts} (4 respondents) sees students describing using AI as a substitute for the domain expertise they lacked and points to a possible integration strategy in which AI is positioned explicitly as a contextualization tool rather than a generation tool. Finally, two respondents proposed an \emph{empirical comparison} approach: having students produce artifacts with and without AI assistance, then comparing and discussing the results in class, to expose the gap between AI capability and course-standard quality and serving both as an AI literacy activity and as an empirical experimentation about requirements engineering in practice. \textbf{Answer to RQ4:} A large majority of the respondents expressed the desire for a better integration of LLMs in the course, principally in the form of classes about efficient prompt engineering techniques, or as a companion tool to support repetitive and verbose tasks in student projects.

\section{Discussion}
\label{sec:disc}

Our findings are largely consistent with prior work on LLM adoption in higher education and RE education, while also revealing several RE-specific characteristics emerging from project-based workflows. Similar to previous studies, our respondents perceived LLMs as highly useful for productivity, brainstorming, and supporting writing-intensive activities, despite simultaneously expressing concerns regarding reliability, correctness, and excessive dependence on AI-generated outputs. In the specific context of RE education, our findings further reinforce the importance of human oversight and critical evaluation in AI-assisted RE tasks, while also finding that students naturally integrated LLMs into complex projects, primarily using them as reviewers, consistency checkers, and contextualization tools rather than as autonomous artifact generators. Several of the failure modes directly reflect the technical limitations highlighted in recent LLM4RE literature reviews. The strong influence of free premium subscriptions on tool adoption and the widespread use of multi-model comparison workflows emerged as distinctive aspects of our study that are currently underexplored in existing Requirements Engineering education literature.

\section{Lessons Learned}
\label{sec:lessons}

In this section, we discuss the main lessons learned from our experience, to provide actionable guidelines for course instructors and researchers in the field:

\begin{itemize}
    \item \textbf{LLMs are a baseline tool within student projects:} All the respondents used at least one LLM during their project activities without any prohibition or encouragement. Course designs that assume no LLM use is done are misaligned with actual student practice and do not guarantee better alignment with the intended learning outcomes. Acknowledging and guiding the use of LLMs is more productive than ignoring or prohibiting it.
    \item \textbf{The highest-valued AI activities in RE are review and brainstorming:} The most satisfactory use of LLMs involved independent artifact production, preliminary brainstorming, or validation, correction and completion of drafts, preserving student cognitive engagement while leveraging LLM agents' capacity for consistency checking. Instructors could formalize the LLM-assisted workflow by designing tasks involving an AI-assisted review phase.
    \item \textbf{Course-specific context is a barrier to useful AI output:} Students are aware that LLMs are not capable of understanding what the course requires, even by using examples and providing additional material. Instructors can address this by publishing course-specific AI prompt templates, or by making the prompt engineering phase an explicit and assessable part of the project.
    \item \textbf{Prompt engineering and critical evaluation should be explicitly taught:} The most common student proposal for LLM integration was to teach them how to write effective prompts and how to evaluate AI output critically. A short dedicated module covering prompt design, output evaluation criteria and hallucination detection strategies, organized early in computer engineering curricula, would provide a skill directly transferable to professional settings. 
    \item \textbf{Access inequality distorts tool choice and potentially learning outcomes:} The dominance of Gemini as the preferred tool was driven primarily by free Pro subscriptions for student accounts, not by perceived quality, while students who had access to paid subscriptions had better-quality AI assistance. This differential is an equity concern that educators and institutions should address through, for example multi-model institutional licences.
\end{itemize}

\section{Threats to Validity}
\label{sec:threats}
 
We report the threats to the validity according to the classification of Feldt and Magazinius \cite{feldt2010validity}. \textbf{Internal validity.} The open coding was performed by a single researcher; a second pass was conducted to improve consistency, but does not substitute formal inter-rater agreement procedures. The survey was administered after all four tasks were performed, introducing recall bias for earlier tasks. \textbf{Construct validity.} Self-reported data on AI use intensity and output quality are subject to social desirability bias: students may have underreported use, or framed it more critically to avoid negative instructor judgment, despite LLM use not being discouraged. We did not collect data on the actual content of prompts or AI outputs, limiting the depth of possible analysis of AI-human interaction. \textbf{External validity.} The study is based on a single class at a single institution, in a specific national and institutional context. Gemini's prevalence reflects a free student subscription programme that may not exist elsewhere. The findings should be treated as exploratory and hypothesis-generating rather than as generalizable claims.
 
\section{Conclusion}
\label{sec:conclusion}

In this paper, we reported the results of a preliminary experience integrating LLMs into the RE phase of a master's Software Engineering course project. We found that students used LLMs extensively, due to the verbose and time-consuming tasks, but also perceived the produced output as inferior to their own work in most cases. Several failure modes were also reported by students, such as style and format mismatches, hallucinated content, over-engineering, and diagram generation limitations; we argue that the reason behind these issues is the gap between the general-purpose capabilities of current LLMs and the course-specific, context-dependent nature of requirements engineering artifacts. The most consistent request from students was support from instructors in the form of guidance on prompt engineering skills, domain knowledge, and critical evaluation capacity; we argue that the most important curricular response to the LLM era in RE education is not the definition of policies but a pedagogical need to provide the aforementioned guidance. This is a form of AI literacy that is both educationally valuable and relevant to professional practice, due to future software engineers being expected to work alongside AI tools. In our future work, we plan to replicate the survey in the second half of the course project to observe the LLMs' impact on testing tasks; furthermore, we plan to adjust our course syllabus to empirically evaluate the impact on the quality of the generated artifacts when students receive AI guidance up-front before working on project activities.

\end{document}